\magnification 1240 \baselineskip=14.5pt \voffset=-0.4truecm
\hoffset=-1.1truecm \vsize=24.truecm \hsize=17.8truecm
\font\gf=cmb10 scaled 1260 \null \vskip 2.5truecm \centerline{\gf
Coupled plasmon-phonon modes in a two-dimensional} \vskip
0.2truecm \centerline{\gf electron gas in the presence of Rashba
effect}

\vskip 1.5truecm \centerline{W. Xu$^{1)}$, M.P. Das$^{1)}$ and
L.B. Lin$^{2)}$}

\vskip 0.5truecm \centerline{$^{1)}$Department of Theoretical
Physics}\vskip 0.1truecm \centerline{Research School of Physical
Sciences and Engineering} \vskip 0.1truecm \centerline{Australian
National University, Canberra, ACT 0200, Australia} \vskip
0.3truecm \centerline{$^{2)}$ Department of Physics, Sichuan
University} \vskip 0.1truecm \centerline{Chengdu - 610064,
Sichuan, People's Republic of China}

 \vskip
1.0truecm

Elementary electronic excitation is studied theoretically for a
two-dimensional electron gas in the presence of spin orbit (SO)
interaction induced by Rashba effect. We find that in such a
system, coupled plasmon-phonon excitation can be achieved via
intra- and inter-SO electronic transitions. As a result, six
branches of the coupled plasmon-phonon oscillations can be
observed. The interesting features of these excitation modes are
analyzed.

\vfill\eject

Progress made in realizing spin polarized electronic systems has
led to recent proposals of the novel electronic devices, such as
spin-transistors [1], spin-waveguides [2], spin-filters [3],
quantum computers [4], etc. One important aspect in the field of
``spintronics'' is to investigate electronic systems with finite
spin splitting at zero magnetic field. It is known that in
semiconductor-based two-dimensional electron gas (2DEG) systems,
the zero-field spin splitting can be realized from inhomogeneous
surface electric field induced by the presence of the
heterojunction. This feature is known as the Rashba effect [5]. In
these systems the strength of the spin-splitting and spin-orbit
interaction (SOI) can be altered by applying a gate voltage [6] or
varying sample growth parameters [7]. At present most of the
published work is focused on electronic and transport properties
of the 2DEGs in the presence of SOI. In order to understand these
novel material systems more deeply and to explore their further
applications to the practical devices, it is necessary for us to
examine the roles which electronic many-body effects and phonons
can play in a 2DEG with SO coupling.

Here we consider an interacting 2DEG where SOI and electron-phonon
(e-p) interaction are present. Our aim is to obtain coupled
plasmon-phonon excitation modes for this system. For a typical
2DEG in the xy-plane in narrow gap semiconductors, such as
InGaAs/InAlAs quantum wells, the noninteracting Schr\"odinger
equation including the lowest order of SOI can be solved
analytically [2]. Applying the electron wavefunctions to the
electron-electron (e-e) interaction Hamiltonian induced by the
Coulomb potential, the space Fourier transform of the matrix
element for bare e-e interaction is written as
$$V_{\alpha\beta}({\bf k},{\bf q})=V_q F_0 (q)
G_{\alpha\beta}({\bf k},{\bf q}).\eqno(1a)$$ Here,
$\alpha=(\sigma', \sigma)$ with $\sigma=\pm 1$ referring to
different SOs, ${\bf k}=(k_x,k_y)$ is the electron wavevector
along the 2D-plane, ${\bf q}=(q_x,q_y)$ is the change of ${\bf k}$
during a scattering event, $V_q=2\pi e^2/\epsilon_\infty q$ with
$\epsilon_\infty$ being the high-frequency dielectric constant,
and $$G_{\alpha\beta}({\bf k},{\bf q})={1+\alpha A_{\bf kq}\over
2}\delta_{\alpha,\beta}+i{\alpha B_{\bf kq}\over
2}(1-\delta_{\alpha,\beta})\eqno(1b)$$ with $A_{\bf kq}=(k+q{\rm
cos}\theta)/|{\bf k}+{\bf q}|$, $B_{\bf kq}=q{\rm sin}\theta/|{\bf
k}+{\bf q}|$, and $\theta$ being the angle between ${\bf k}$ and
${\bf q}$. Furthermore, the space Fourier transform of the matrix
element for bare e-p interaction can be written as
$$V_\alpha^{ph}({\bf k},{\bf q};\Omega)=\sum_{q_z}D_0 (\omega_Q,
\Omega)|U_\alpha ({\bf k},{\bf Q})|^2, \eqno(2)$$ where ${\bf
Q}=({\bf q},q_z)$ is the phonon wavevector, $\omega_Q$ is the
phonon frequency, $D_0(\omega_Q,\Omega)=2\hbar
\omega_Q/[(\hbar\Omega)^2-(\hbar\omega_Q)^2]$ is the bare phonon
propagator, $|U_\alpha ({\bf k},{\bf q})|^2=|W_{\bf Q}|^2G_0(q_z)
A_\alpha ({\bf k},{\bf q})$ is the square of the e-p interaction
matrix element, $A_\alpha ({\bf k},{\bf q})=(1+\alpha A_{\bf
kq})/2$ is a spin-dependent element, and $W_{\bf Q}$ is the e-p
coupling coefficient. It should be noted that in contrast to a
conventional 2DEG (C2DEG) for which the bare e-e and bare e-p
interactions do not depend on ${\bf k}$ [8], $V_{\alpha\beta}
({\bf k},{\bf q})$ and $V_\alpha^{ph} ({\bf k},{\bf q};\Omega)$
for a 2DEG with SOI depend not only on ${\bf q}$ but also on {\bf
k}, because the spin splitting depends explicitly on {\bf k}. In
the present study, we consider the case of a narrow quantum well
in which only one electronic subband is present. Thus,
$F_0(q)=\int dz_1 \int dz_2\ |\psi_0(z_1)|^2 |\psi_0 (z_2)|^2 {\rm
exp}(-q|z_1-z_2|)$ and $G_0(q_z)=|<0|e^{iq_zz}|0>|^2$ with
$|0>=\psi_0 (z)$ being the electron wavefunction along the growth
direction.

From electron energy spectrum obtained by solving the
Schr\"odinger equation, we derive the retarded and advanced
Green's functions for electrons. Using these Green's functions,
$V_{\alpha\beta}({\bf k},{\bf q})$ and $V_\alpha^{ph}({\bf k},{\bf
q};\Omega)$ in a diagrammatic self-consistent theory [9] (also see
Fig. 1), the effective e-e interaction is given by
$$V_{eff}({\bf k},{\bf q};\Omega)=[V_{\alpha\beta}({\bf k},{\bf
q})+V_\alpha^{ph}({\bf k},{\bf
q};\Omega)]\epsilon_{\alpha\beta}^{-1} ({\bf k},{\bf
q};\Omega).\eqno(3)$$ Here,
$$\epsilon_{\alpha\beta}({\bf k},{\bf q};\Omega)=\delta_{\alpha,
\beta}\delta({\bf k})-[V_{\alpha\beta}({\bf k},{\bf q})+
V_\alpha^{ph}({\bf k},{\bf q};\Omega)]\Pi_\beta ({\bf k},{\bf q};
\Omega) \eqno(4)$$ is the dielectric function matrix element and
$$\Pi_{\sigma'\sigma}({\bf k},{\bf q};\Omega )={f[E_{\sigma'}
({\bf k}+{\bf q})]-f[E_{\sigma}({\bf k})]
\over\hbar\Omega+E_{\sigma'}({\bf k}+{\bf q})-E_{\sigma}({\bf
k})+i\delta} \eqno(5)$$ is the pair bubble in the absence of e-e
coupling with $f(E)$ being the Fermi-Dirac function. In Eq. (5),
$$E_\sigma ({\bf k})=E_\sigma (k)=\hbar^2 k^2/2m^*+\sigma\alpha_R
k \eqno(6)$$ is the energy spectrum of a 2DEG in the presence of
SOI, with $m^*$ being the electron effective mass and $\alpha_R$
the Rashba parameter which measures the strength of the SOI.

For a 2DEG with SOI, the effective e-e interaction and dielectric
function matrix depend not only on {\bf q} but also on {\bf k}, in
contrast to a C2DEG. After summing $\epsilon_{\alpha\beta}({\bf
k},{\bf q};\Omega)$ over {\bf k} and setting $1=(++)$, $2=(+-)$,
$3=(-+)$ and $4=(--)$, the dielectric function matrix for a 2DEG
with Rashba spin splitting in the presence of e-p scattering is
obtained as $$\epsilon =\left( \matrix{ 1+a_1+b_1 & 0 & 0 & a_4\cr
0 & 1+a_2+b_2 & a_3 & 0 \cr 0 & a_2 & 1+a_3+b_3 & 0 \cr a_1 & 0 &
0 & 1+a_4+b_4} \right).\eqno(7)$$ In Eq. (7), $a_j=-[V_q
F_0(q)/2]B_j({\bf q},\Omega)$ and $b_j=-\sum_{q_z}D_0
(\omega_Q,\Omega) |W_{\bf Q}|^2 G_0(q_z) B_j ({\bf q},\Omega)$ are
induced respectively by e-e and e-p interaction. $B_j({\bf
q},\Omega)= \sum_{\bf k} (1\pm A_{\bf kq})\Pi_j ({\bf k},{\bf
q};\Omega)$, where upper (lower) case refers to $j=1$ or $4$ for
intra-SO transitions ($j=2$ or $3$ for inter-SO transitions). The
determinant of the dielectric function matrix is then given by
$$|\epsilon|=[(1+a_1+b_1)(1+a_4+b_4)-a_1a_4]
[(1+a_2+b_2)(1+a_3+b_3)-a_2a_3]\eqno(8)$$ which results from
intra- and inter-SO electronic transitions. Thus, the modes of
coupled plasmon-phonon excitation are determined by ${\rm
Re}|\epsilon|\to 0$.

In the present study, we consider an InGaAs-based 2DEG in which
electrons interact strongly with longitudinal optical (LO) phonons
through the F\"rohlich coupling. For electron interaction with
LO-phonons, $\omega_Q\to\omega_{LO}$ the LO-phonon frequency at
long-wavelength limit, $|W_{\bf Q}|^2=2\pi e^2\hbar\omega_{LO}
(\epsilon_\infty^{-1}-\epsilon_s^{-1})/Q^2$ with $\epsilon_s$ and
$\epsilon_\infty$ being respectively the static and high-frequency
dielectric constants. At a long-wavelength (i.e., $q\ll 1$) and a
low-temperature (i.e., $T\to 0$) limit, we have $${\rm
Re}|\epsilon|\simeq \Bigl[1-{\omega_p^2\over\Omega^2}{
\Omega^2-\omega_{TO}^2\over \Omega^2-\omega_{LO}^2}\Bigl(
1-{\omega_--\omega_+ \over\omega_0/2}\Bigr)\Bigr]
\Bigl[1-{\omega_p^2\over\omega_0\Omega}{\Omega^2-\omega_{TO}^2
\over \Omega^2-\omega_{LO}^2}{\rm ln} \Bigl({\Omega+\omega_-
\over\Omega-\omega_-}{\Omega-\omega_+\over\Omega+\omega_+}
\Bigr)\Bigr].\eqno(9)$$ Here, the first (second) term on the
right-hand side is induced by intra-SO (inter-SO) transitions,
$\omega_{TO}=\sqrt{\epsilon_\infty/\epsilon_s}\omega_{LO}$ is the
TO-phonon frequency, $\omega_\pm=4\alpha_R\sqrt{\pi n_\pm}/\hbar$
with $n_\pm$ being the electron density in the $\pm$ spin channel,
$\omega_0=16\pi n_e\hbar/m^*$, and $\omega_p=(2\pi e^2 n_e
q/\epsilon_\infty m^*)^{1/2}$ is the plasmon frequency of a 2DEG
in the absence of SOI with $n_e=n_++n_-$ being the total electron
density of the system. Moreover, it can be shown that at
low-temperature limit, electron density in different SOs is
$$n_\pm=(n_e/ 2)\mp (k_\alpha/ 2\pi)\sqrt{2\pi n_e -k_\alpha^2}
\eqno(10)$$ for case of $n_e>k_\alpha^2/\pi$ with $k_\alpha=m^*
\alpha_R/\hbar^2$. When $n_e\le k_\alpha^2/\pi$, only spin-down
states are occupied by electrons and, therefore, $n_+=0$ and
$n_-=n_e$.

In the presence of SOI, the collective excitation from a 2DEG can
be achieved via electron transitions in different spin channels.
From Eq. (9), we see that the coupled plasmon-phonon frequency
induced by intra-SO excitation is given by $$\Omega_+=\omega_{LO}
+{a\over 2}\Bigl(1-{\epsilon_\infty\over\epsilon_s}\Bigr)
{\omega_p^2\over\omega_{LO}} \ \ \ \ \ {\rm and} \ \ \ \ \
\Omega_-=\omega_p\sqrt{a{\epsilon_\infty\over
\epsilon_s}},\eqno(11)$$ where
$a=1-2(\omega_--\omega_+)/\omega_0$; those induced by inter-SO
transitions can be obtained by solving
$${\rm ln}\Bigr({\Omega+\omega_-\over\Omega-\omega_-}\cdot
{\Omega-\omega_+\over\Omega+\omega_+}\Bigl)={\omega_0\Omega
\over\omega_p^2}{\Omega^2-\omega_{LO}^2\over\Omega^2-
\omega_{TO}^2}.\eqno(12)$$ Thus, two (four) branches of the
coupled plasmon-phonon excitation can be observed via intra-SO
(inter-SO) electronic transitions.

Now we present the results of our calculations for InGaAs-based
quantum well structures. The material parameters are known [10]:

\item{1)} the electron effective mass $m^*=0.042 m_e$ with $m_e$
being the rest-electron mass;

\item{2)} the high-frequency and static dielectric constants are
respectively $\epsilon_\infty=12.3$ and $\epsilon_s=14.6$; and

\item{3)} the LO-phonon energy $\hbar\omega_{LO}=30.9$ meV.

The dependence of coupled plasmon-phonon frequencies induced by
intra- (in (a)) and inter-SO (in (b)) excitation on
$\omega_p=(2\pi e^2 n_e q/\epsilon_\infty)^{1/2}$ or $q$, on
Rashba parameter $\alpha_R$ and on total electron density $n_e$
are shown respectively in Figs. 2 - 4. From these results and from
Eqs. (11) and (12), it can be found that at a long-wavelength
limit, excitations with frequency about $\omega_{LO}$ can be
generated via both intra- ($\Omega_+\sim \omega_{LO}$) and
inter-SO ($\Omega_4\sim \omega_{LO}$) transitions and they depend
very weakly on $q$ and sample parameters (such as $\alpha_R$ and
$n_e$). Another mode induced by intra-SO transition,
$\Omega_-\sim\omega_p\sim q^{1/2}$, is acoustic-like and depends
weakly on $\alpha_R$ and $n_e$. The excitation with frequency
about $\Omega_3\sim\omega_{TO}$ can only be generated via inter-SO
transitions and its dependance on $q$, $\alpha_R$ and $n_e$ are
negligible, which implies that the TO-phonon mode can be excited
via inter-SO transitions in a InGaAs-based 2DEG. Although
$\Omega_1$ and $\Omega_2$ induced by inter-SO transitions should,
in principle, depend on $q$ via $\omega_p$ (see Eq. (12)), the
numerical results shown in Fig. 2 suggest that at long-wavelength
limit, over a wide regime of $\omega_p$ or $q$,
$\Omega_1\to\omega_+=4\alpha_R\sqrt{\pi n_+}/\hbar$ and $\Omega_2
\to\omega_-=4\alpha_R\sqrt{\pi n_-}/\hbar$ depend very little on
$q$ and can differ significantly from the phonon frequencies and
from $\omega_p$. This indicates that $\Omega_1$ and $\Omega_2$ are
optic-like and they rely very strongly on sample parameters such
as $\alpha_R$ and $n_e$ (see Figs. 3 and 4).

 The important conclusions drawn from this work are, for a 2DEG
with SOI,

\item{(1)} LO-phonon excitation can be achieved via intra- and
inter-SO electronic transitions;

\item{(2)} TO-phonon mode can only be generated via inter-SO
excitation; and

\item{(3)} $\Omega_1$ and $\Omega_2$ induced by inter-SO
excitations are optic-like and very sensitive to sample
parameters.

There are many investigations on the coupled plasmon-phonon modes
in semiconductor-based 2DEG systems without including SOI [9,11].
In comparing these results with those obtained in our present
study (where SOI is present), we note some interesting features in
the collective excitation modes, especially those excited through
inter-SO transitions. It may be noted that the state-of-the-art
material engineering and micro- and nano-fabrication techniques
have made it possible to achieve an experimentally observable
Rashba spin-splitting in, e.g., InGaAs-based 2DEG systems. Very
recent experimental results [12] in this system show that the
Rashba parameter $\alpha_R$ can reach up to $3\sim 4 \times
10^{-11}$ eVm. Our results shown in Figs. 2 - 4 indicate that when
$\alpha_R\geq 10^{-11}$ eVm, a significant separation between
$\Omega_{1,2}$ induced by inter-SO transition can be achieved.

At present, magneto-transport measurement is a powerful and most
popularly used experimental technique to identify Rashba
spin-splitting in a 2DEG [6,7]. Using this technique to determine
the Rashba parameter and electron density in different spin
branches, one needs to apply high magnetic fields at
low-temperatures so that Shubnikov-de Hass oscillations are
observed. From our investigations we suggest that the Rashba spin
splitting can be probed by optical measurements. It should be
noted that the dispersion of the coupled plasmon-phonon energies
(or frequencies) is of the same order of magnitude as energy
splitting in different spin branches. Together with a fact that it
is not so easy to measure the dispersion relation of the coupled
plasmon-phonon modes in a 2DEG even without inclusion of SOI, we
think it may be difficult to measure optically the dispersion
relation of the coupled plasmon-phonon mode in a 2DEG with SOI. In
the absence of SOI, the dispersion of the coupled plasmon-phonon
mode can only be determined by using techniques, such as grating
couplers [13]. However, Raman scattering [14] and ultrafast
pump-and-probe experiments [15] have been carried out recently to
study coupled plasmon-phonon modes in semiconductor-based 2DEG
systems without SOI. These optical experiments conveniently
measure optic-like excitation modes. Thus, we propose here coupled
plasmon-phonon modes in a spin split 2DEG to be detected optically
because most of them are optic-like. In particular, if we can
measure $\omega_\pm=4\alpha_R\sqrt{\pi n_\pm}/\hbar$ (the
magnitude of the frequencies and their separation are of the order
of THz), we can determine the Rashba parameter and total electron
density of the system. We see from Figs. 2 - 4 that when
$\alpha_R\geq 10^{-11}$ eVm, $\Omega_2-\Omega_1$ is much larger
than $\Omega_4-\Omega_3\simeq \omega_{LO}-\omega_{TO}$. Because
even $\omega_{LO}$ and $\omega_{TO}$ in InGaAs can be resolved in,
e.g., Raman spectra [10], $\Omega_{1,2}$ induced by inter-SO
transition can be more easily resolved in an optical experiment.
Finally we suggest these theoretical predictions require
experimental verification.

\vskip 0.5truecm \noindent {\bf Acknowledgment:} One of us (W. X.)
is a Research Fellow of the Australian Research Council. This work
is also partly supported by the National Natural Science
Foundation of China.

\vfill\eject \centerline{\bf References} \vskip 0.5truecm

\item{[1]} B. Datta and S. Das, Appl. Phys. Lett. {\bf 56},
665 (1990).
\item{[2]} X.F. Wang, P. Vasilopoulos, and F.M. Peeters, Phys.
Rev. {\bf B 65}, 165217 (2002).

\item{[3]} T. Koga, J. Nitta, H. Takayanagi, and S. Datta, Phys.
Rev. Lett. {\bf 88}, 126601 (2002).

\item{[4]} Y. Ohno, D.K. Young, B. Beschoten, F. Matsukura, H.
Ohno, and D.D. Awschalom, Nature {\bf 402}, 790 (1999).

\item{[5]} E.I. Rashba and V.I. Sheka, in {\it Landau
Level Spectroscopy}, edited by G. Landwehr and E.I. Rashba
(North-Holland, Amsterdam, 1991).

\item{[6]} J. Nitta, T. Akazaki, H. Takayanagi, and T. Enoki,
Phys. Rev. Lett. {\bf 78}, 1335 (1997).

\item{[7]} J. Luo, H. Munekata, F.F. Fang, and P.J. Stiles, Phys.
Rev. {\bf B 41}, 7685 (1990).

\item{[8]} See, e.g., T. Ando, A.B. Fowler, and F. Stern, Rev.
Mod. Phys. {\bf 54}, 437 (1982).

\item{[9]} See, e.g., R.Jalabert and S. Das Sarma, Phys. Rev.
{\bf B 40}, 9723 (1989).

\item{[10]} See, e.g., S. Adachi, {\it Physical Properties of
III-V Semiconductor Compounds} (John Wiley \& Sons, NY, 1992).

\item{[11]} See, e.g., F.M. Peeters, X.G. Wu, and J.T. Devreese,
Phys. Rev. {\bf B 36}, 7518 (1987); L. Wendler and R. Pechstedt,
J.Phys.: Condens. Matter {\bf 2}, 8881 (1990); W.H. Backes, F.M.
Peeters, F. Brosens, and J.T. Devreese, Phys. Rev. {\bf B 45},
8437 (1992); M. Reizer, {\it ibid} {\bf B 61}, 40 (2000).

\item{[12]} Y. Sato, T. Kita, S. Gozu, and S. Yamada, J. Appl.
Phys. {\bf 89}, 8017 (2001); D. Grundler, Phys. Rev. Lett. {\bf
84}, 6074 (2000).

\item{[13]} See, e.g., M. Vo$\beta$eb\"urger, H.G. Roskos, F.
Wolter, C. Waschke, and H. Kurz, J. Opt. Soc. Am {\bf B 13}, 1045
(1996).

\item{[14]} See, e.g., P. Brockmann, J.F. Young, P. Hawrylak, and
H.M. van Driel, Phys. Rev. {\bf B 48}, 11423 (1993).

\item{[15]} See, e.g., T. Dekorsy, A.M.T. Kim, G.C. Cho, H. Kurz,
A.V. Kuznetsov, and A. F\"orster, Phys. Rev. {\bf B 53}, 1531
(1996).

\vfill\eject \centerline{\bf FIGURE CAPTIONS} \vskip 0.5truecm
\noindent {\bf Fig. 1.} Effective e-e interaction
(double-solid-lines) in the presence of phonon scattering. Here,
the dashed-line is the bare e-e interaction, the dotted-line is
induced by electron-phonon scattering, and the bubble refers to
the bare pair-bubble.

\vskip 0.5truecm \noindent {\bf Fig. 2.} Dependence of the coupled
plasmon-phonon frequency induced by intra- (in (a)) and inter-SO
(in (b)) excitation on $\omega_p=(2\pi e^2 n_e q/\epsilon_\infty
m^*)^{1/2}$ for fixed Rashba parameter $\alpha_R$ and total
electron density $n_e$. Here, $\omega_{LO}$ and $\omega_{TO}$ are
respectively the LO- and TO-phonon frequencies and
$\omega_\pm=4\alpha_R\sqrt{\pi n_\pm}/\hbar$.

\vskip 0.5truecm \noindent {\bf Fig. 3.} Coupled plasmon-phonon
frequency induced by intra- (in (a)) and inter-SO (in (b))
transitions as a function of Rashba parameter for fixed $\omega_p$
and $n_e$ as indicated.

\vskip 0.5truecm \noindent {\bf Fig. 4.} Coupled plasmon-phonon
frequency caused by intra- (in (a)) and inter-SO (in (b))
excitations versus total electron density for fixed $\alpha_R$ and
$\omega_p$.

\vfill\eject \null \topinsert \input psfig.sty
\centerline{\psfig{figure=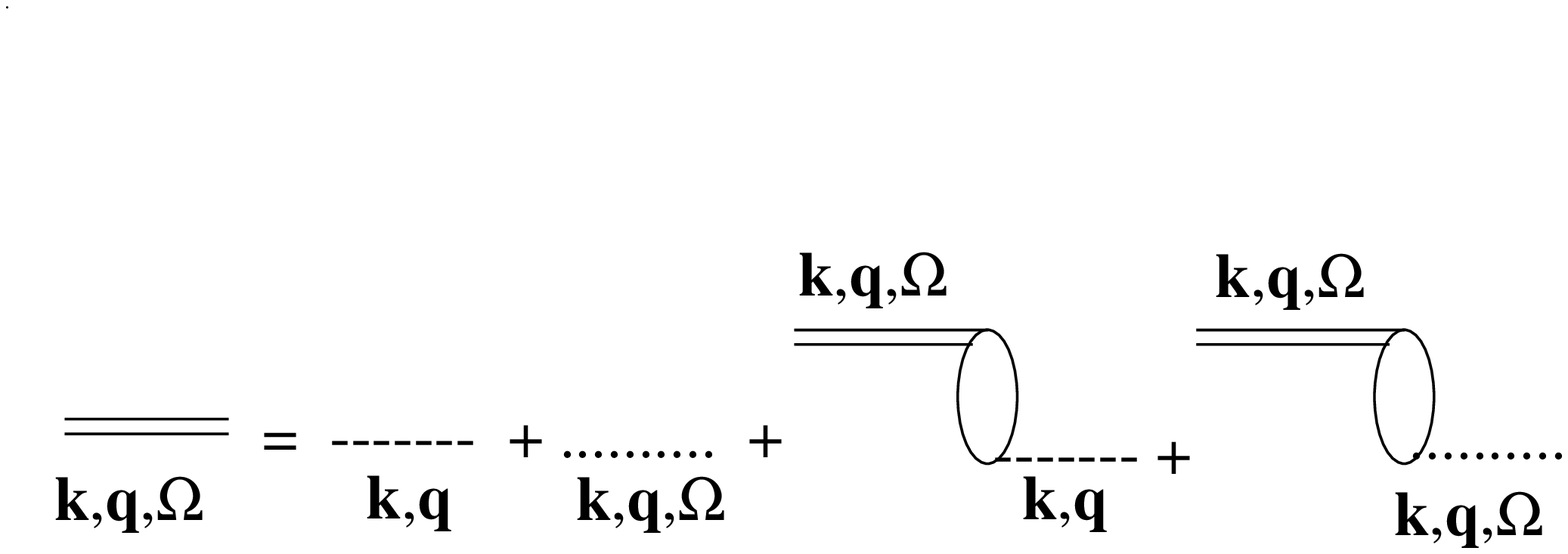,width=14truecm,angle=0}} \vskip
2.0truecm \noindent {\bf W. Xu et. al. Figure 1.} \endinsert

\vfill\eject \null \topinsert \input psfig.sty
\centerline{\psfig{figure=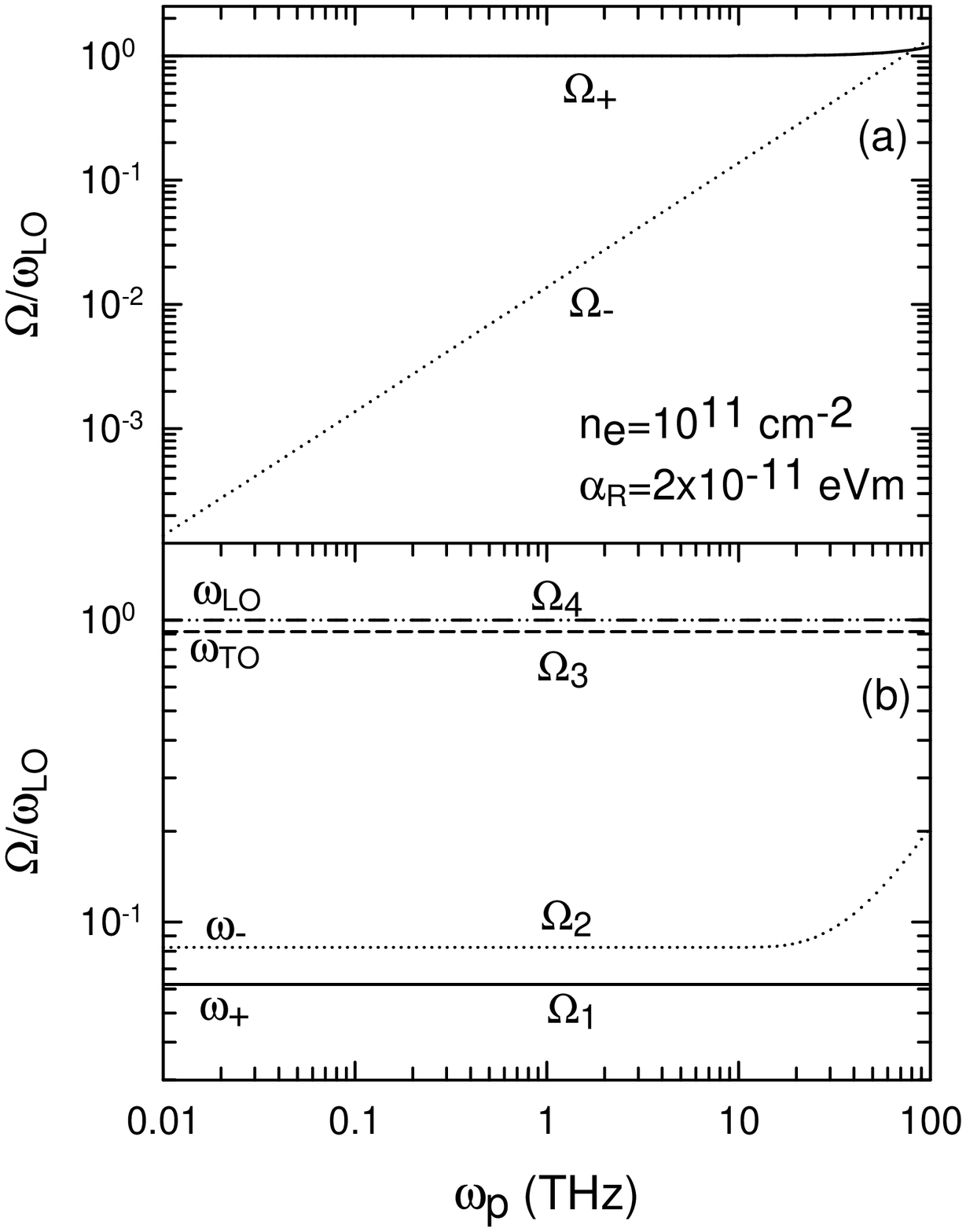,width=14truecm,angle=0}} \vskip
2.0truecm \noindent {\bf W. Xu et. al. Figure 2.}
\endinsert

\vfill\eject \null \topinsert \input psfig.sty
\centerline{\psfig{figure=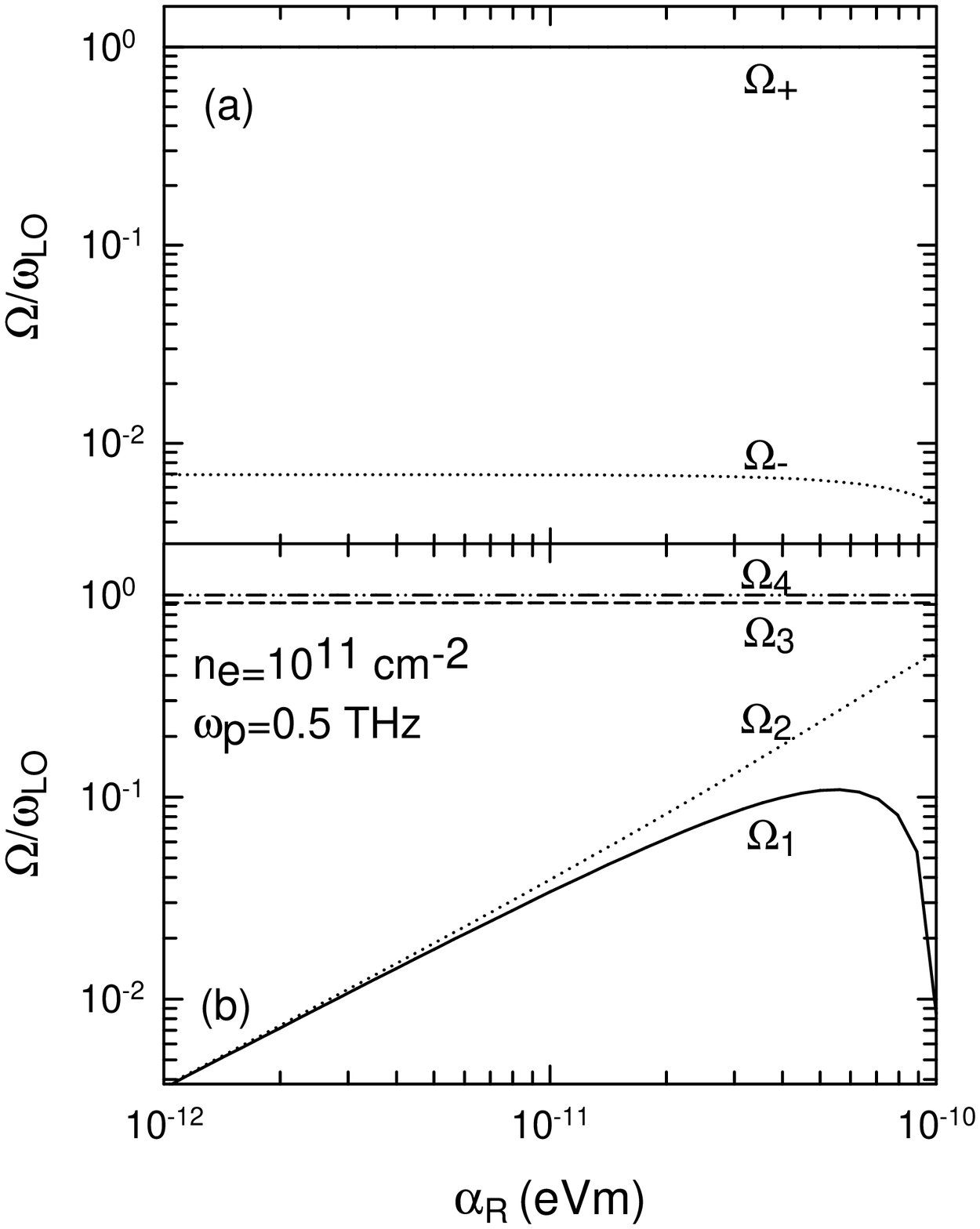,width=14truecm,angle=0}} \vskip
2.0truecm \noindent {\bf W. Xu et. al. Figure 3.}
\endinsert

\vfill\eject \null \topinsert \input psfig.sty
\centerline{\psfig{figure=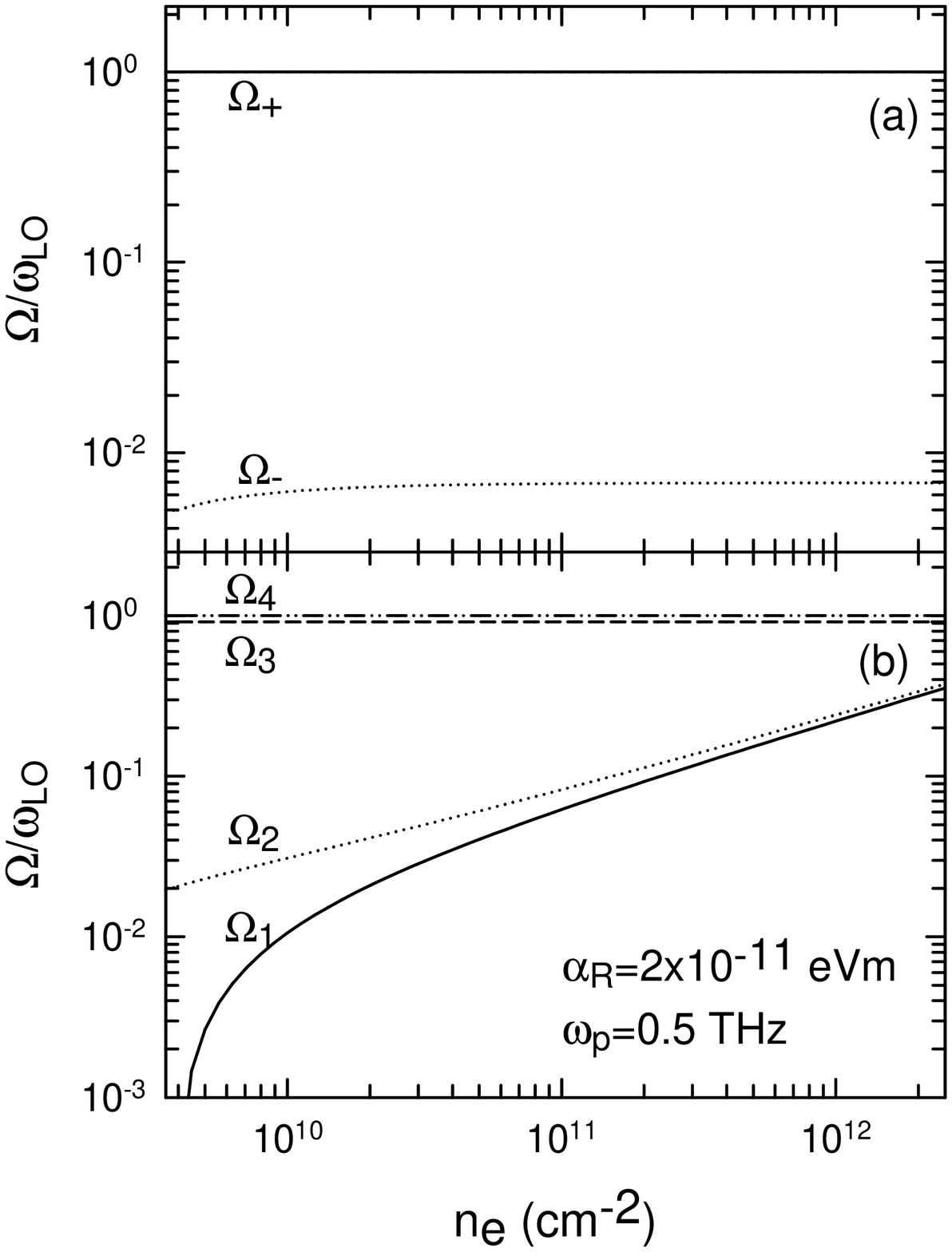,width=14truecm,angle=0}} \vskip
2.0truecm \noindent {\bf W. Xu et. al. Figure 4.} \endinsert

\end